\documentclass[prl,twocolumn,showpacs,amsmath,amssymb,floats,floatfix,prl,aps,superscriptaddress,showkeys]{revtex4}
\usepackage{subfig}
\usepackage{graphicx}
\usepackage{epsfig}
\usepackage{dcolumn}
\usepackage{bm}
\usepackage{times}
\usepackage{soul}
\usepackage{color}

\graphicspath{{C:/Users/Gabriel/Desktop/figures/}}
\begin{document}


\title{Berry-phase in a periodically driven single molecule magnet transistor}

\author{Gabriel Gonz\'alez}

\email{gabriel.gonzalez@uaslp.mx}
\affiliation{C\'atedra CONCAYT, Universidad Aut\'onoma de San Luis Potos\'i, San Luis Potos\'i, 78000 MEXICO}
\affiliation{Coordinaci\'on para la Innovaci\'on y la Aplicaci\'on de la Ciencia y la Tecnolog\'ia, Universidad Aut\'onoma de San Luis Potos\'i,San Luis Potos\'i, 78000 MEXICO}

\date{\today}

\begin{abstract}
We consider the electron transport through a single molecule magnet transistor in the presence of a local transverse magnetic field and ac-driven gate voltage. We calculate the conductance as a function of the electron energy and transverse magnetic field by using the Floquet and Landauer formalism. We show that the time periodic potential causes zero transmission resonances that oscillate as a function of the transverse magnetic field due to the Berry phase interference associated with two quantum tunneling paths. We find that these Berry phase oscillations can be detected in the conductance as a function of the transverse magnetic field for an incoming electron with a specific energy.
\end{abstract}



\maketitle

\begin{figure}[ht]
\includegraphics[width=8cm]{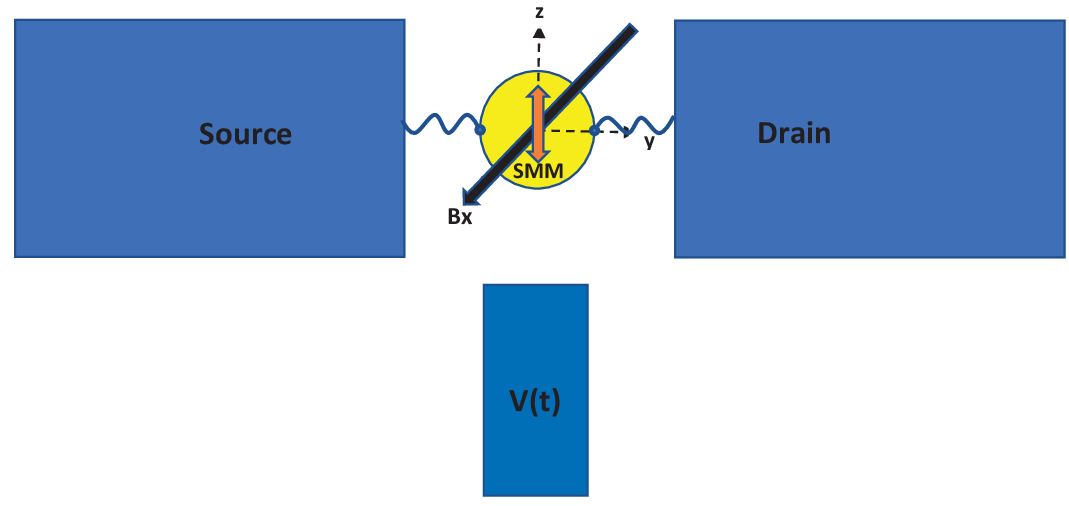}
\caption{Schematic illustration of a single molecule magnet transistor attached to
two leads and controlled by a time dependent gate
voltage and a transverse magnetic field. }
\label{fig:setup}
\end{figure}

Considerable efforts have been devoted to the constant miniaturization of electronic devices in the semiconductor-based industry. The idea to use single molecules as electronic components appeared for the first in the 1974,\cite{Avi} but it had to wait until the appearance of nanoscience which enable individual molecules to be connected between two contact leads.\cite{Rei} Recently, a number of experimental studies haven been conducted to study electron transport through a single molecule magnet.\cite{delBarco}
Single molecule magnets (SMM), such as Mn$_{12}$ (see Ref. \onlinecite{experiments}) and Fe$_8$ (see
Refs. \onlinecite{Sangregorio}), have been a topic of research of growing interest in the nanoscience since experiments in bulk samples provided for the first time evidence of quantum spin tunneling at low temperatures and Berry-phase interference effects in the presence of a transverse magnetic field which leads to a vanishing of the energy splitting.\cite{garg,garg1} Remarkably, a transverse magnetic field can be tuned to topologically quench the two resonant spin states of a SMM.\cite{LossDelftGarg,Delft,Zener,Gunther,Leuenberger_Berry} Electron transport through a SMM in a three terminal configuration have already been studied in several experiments where they have shown the Coulomb blockade. Oscillations of the Kondo effect as a function of the transverse magnetic field were also investigated.\cite{future,gabriel}  Theoretical evidence of the Berry-phase interference effect in electron transport in a SMM transistor with opposite spin polarized leads have also been studied.\cite{gabriel1} On the theoretical side, the electron transport properties in a SMM transistor were mostly investigated within a constant gate voltage. More recently, the problem of electron transport through a periodically driven ferromagnetic quantum barrier with a local magnetic field and ac-driven potential was solved using the Floquet formalism.\cite{eggert}\\ In this letter we present a theoretical study of the Berry-phase effect in electron transport through a SMM transistor under a local time periodic gate voltage and a transverse magnetic field, and show that the time periodic potential causes zero transmission resonances that oscillate as a function of the transverse magnetic field due to the Berry phase interference associated with two quantum tunneling paths. We show that the Berry phase oscillations can be detected in the conductance of the SMM transistor.
In particular, we study the electron transmission through a SMM connected to two contacts with a local transverse magnetic field and a oscillating gate voltage as illustrated in Fig. \ref{fig:setup}. 

We consider transport of unpolarized incoming electrons modeled by a one dimensional tight-binding chain with nearest neighbor hopping amplitude $J$. Taking the on-site energy equal to zero and $\hbar=1$, the effective Hamiltonian reads 
\begin{equation}
\label{eq:Htotal}
{\cal H}_{\rm leads}=-J\sum_{i>0(i<0)}\sum_{\sigma}\left(c^{\dagger}_{i,\sigma}c_{i+1,\sigma}+c^{\dagger}_{i+1,\sigma}c_{i,\sigma}\right)
\end{equation}
where $c^{\dagger}_{i,\sigma}$ and $c_{i,\sigma}$ are the creation and annihilation operators at the $i$th site with spin $\sigma$, respectively.\\
The simplest effective Hamiltonian which describes the Berry-phase of the spin quantum tunneling in SMMs is the so-called giant spin approximation (GSA) model,\cite{romero} and with a local periodically driven gate voltage with angular frequency $\omega$ and amplitude $V_g$ we can write the AC-GSA Hamiltonian as
\begin{equation}
\label{eq:HSMM}
{\cal H}_{\rm SMM}={\cal H}_{\rm GSA}+\sum_{\sigma}\left(\epsilon_{0L}-\!V_g\cos(\omega t)\right)c^{\dagger}_{0,\sigma}c_{0,\sigma}+Un_{0,\uparrow}n_{0,\downarrow},
\end{equation}
where $\epsilon_{0L}$ represents the energy of the orbital level of the SMM which can get tuned by the gate voltage $V_g$, $U$ is the Coulomb energy of the two electrons in the SMM and
\begin{align}
{\cal H}_{\rm GSA}\!\!\!&=&\!\!\!-K_zS_z^2\!-\!B_xS_x\!+\!K(S_x^2\!-\!S_y^2)\!+\!J_{\parallel}S_z(n_{0,\uparrow}-n_{0,\downarrow}) \nonumber \\
& & +J_{\perp}(S^{+}c^{\dagger}_{0,\downarrow}c_{0,\uparrow}+S^{-}c^{\dagger}_{0,\uparrow}c_{0,\downarrow}),
\label{eq:GSA}
\end{align}
where $K_z$ is the uniaxial anisotropy constant, $B_x$ is the local transverse magnetic field, $K$ is the in-plane transverse anisotropy constant, $S_x$, $S_y$ and $S_z$ are quantum molecule spin operators,$n_{0,\sigma}=c^{\dagger}_{0,\sigma}c_{0,\sigma}$ is the number operator, $S^{\pm}=S_{x}\pm iS^{y}$ are spin-flip operators and $J_{\parallel}$ and $J_{\perp}$ are the longitudinal and transverse exchange interactions. \cite{gabriel,timm,misio}
We shall investigate the electron transport for the strong Coulomb regime such that the two-electron configuration do not contribute to transport and we have only two possible states in the SMM which are the uncharged state or the singly charged state. The uncharged or singly charge state of the SMM will give rise to a integer or half integer spin configuration in the molecule. We will consider that the exchange coupling between the electron and the SMM is highly anisotropic so that the spin-flip between the electron and the SMM is suppressed, which means that $J_{\perp}=0$.\\
The ground state energy splitting induced by the quantum tunneling of the spin can be obtained by applying perturbation theory to the GSA Hamiltonian, which is given by \cite{kriz,garg2,kriz1}
\begin{equation}
\label{eq:ts}
\Delta E=\frac{4s}{2^{2s}(2s-1)!K_z^{2s-1}}\prod_{n=1}^{2s}\left(B_x-(2s+1-2n)B_a\right),
\end{equation}
where $B_x$ and $K$ represents the two perturbations, $s$ represents the ground spin state of the SMM and $B_a=\sqrt{2KK_z}$. From Eq. (\ref{eq:ts}) we see that the tunnel splitting is zero when $B_x^{(n)}=(2s+1-2n)B_a$ for $n=1,2,\ldots,2s$. The vanishing of the tunnel splitting for certain values of the transverse magnetic field is due to the destructive inteference between different spin tunneling paths.\\

\begin{figure}[ht]
\includegraphics[width=8cm]{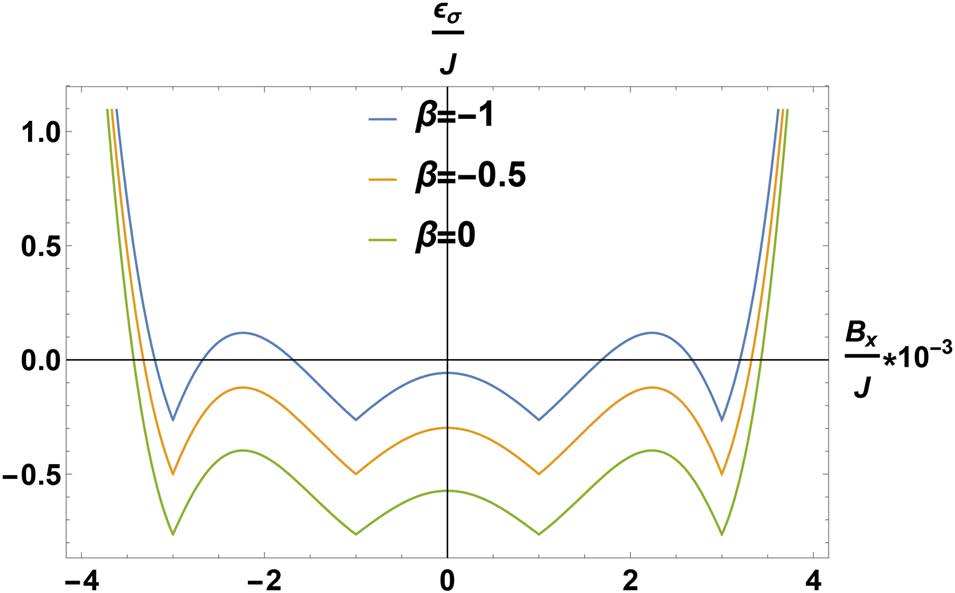}
\caption{Oscillations of the zero transmission resonances as a function of the transverse magnetic field for different values of $\beta$. We used the following values $K_z=10^4 J$, $K=50 J$, $\omega=3 J$, $\epsilon_{OL}=0.5 J$ and $J_{\parallel}=0.75 J$ for our calculations. Note that it is necessary that the uniaxial anisotropy constant $K_z$ be large in order to have oscillations with large amplitudes.}
\label{oscillate}
\end{figure}

\begin{figure}[ht]
\includegraphics[width=8cm]{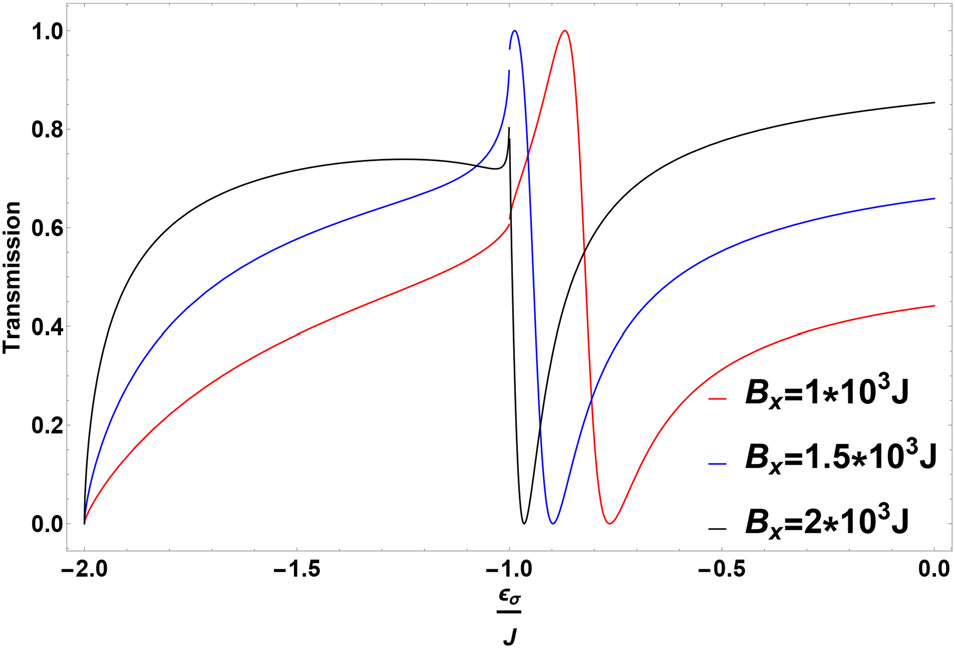}
    \caption{The graph shows the electron transmission in a SMM transistor with homogeneous hopping amplitude for various magnetic field strengths as a function of $\epsilon_{\sigma}$. We used the following values $V_g=1 J$, $K_z=10^4 J$, $K=50 J$, $\omega=3 J$, $\epsilon_{OL}=0.5 J$, $J_{\parallel}=0.75 J$ and $\beta=0$.}
	\label{tbz}
\end{figure}

\begin{figure}[ht]
\includegraphics[width=8cm]{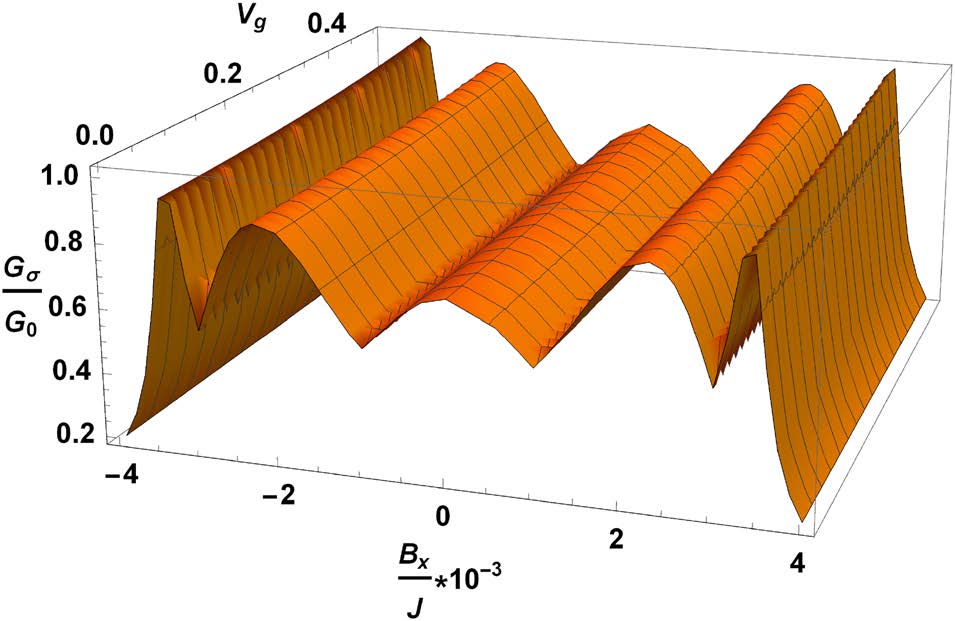}
    \caption{The graph shows the oscillations of the conductance as a function of the transverse magnetic field $B_x$ for a SMM with $s=2$ due to the Berry phase. We used the following values $V_g=1 J$, $\epsilon_{\sigma}=1 J$, $K_z=10^4 J$, $K=50 J$, $\omega=3 J$, $\epsilon_{OL}=0.5 J$, $J_{\parallel}=0.75 J$ and $\beta=0$.}
	\label{conduc}
\end{figure}

\begin{figure}[ht]
\includegraphics[width=8cm]{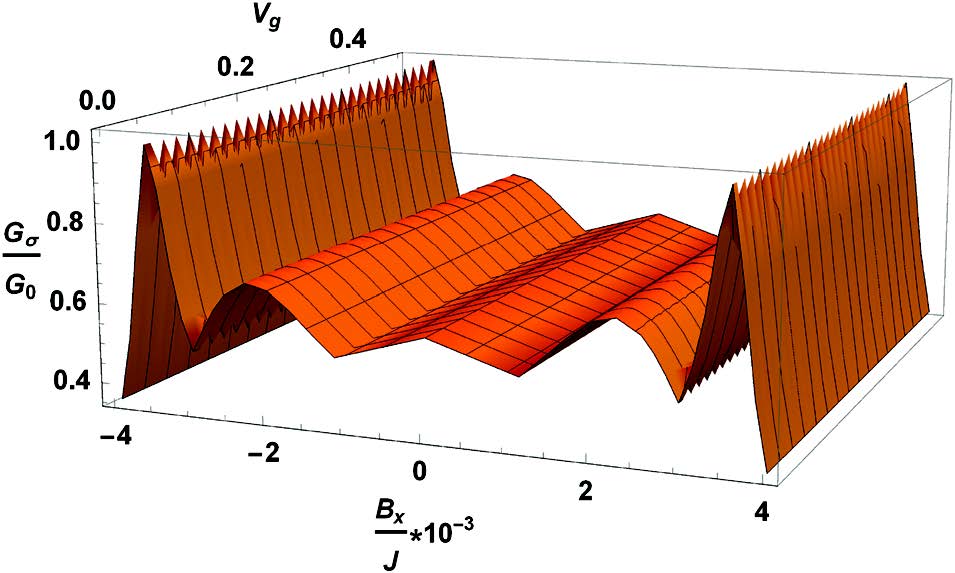}
    \caption{The graph shows the oscillations of the conductance as a function of the transverse magnetic field $B_x$ for a SMM with $s=2$ due to the Berry phase. We used the following values $V_g=1 J$, $\epsilon_{\sigma}=1 J$, $K_z=10^4 J$, $K=50 J$, $\omega=3 J$, $\epsilon_{OL}=0.5 J$, $J_{\parallel}=0.75 J$ and $\beta=-0.5$.}
	\label{conduc1}
\end{figure}

The interaction between the SMM and the leads is given by the following Hamiltonian
\begin{eqnarray}
{\cal H}_{\rm SMM-lead}\!\!\!&=&\!\!\!-J^{\prime}\!\sum_{\sigma}\!\left( 
|s\rangle\langle s+\sigma| c^{\dagger}_{1,\sigma}c_{0,\sigma}\!+\! c^{\dagger}_{0,\sigma}c_{1,\sigma}  |s+\sigma\rangle\langle s| \right. \nonumber \\
& & \!\!\!\!\left.+|s+\sigma\rangle\langle s | c^{\dagger}_{0,\sigma}c_{-1,\sigma} + c^{\dagger}_{-1,\sigma}c_{0,\sigma}|s\rangle\langle s+\sigma |\right)   
\label{eq:SMM-lead}
\end{eqnarray}
where $J^{\prime}$ is the lead-molecule tunneling amplitude. \\
The conductance of the SMM transistor at zero temperature can be obtained using the Landauer formula \cite{landauer}
\begin{equation}
G_{\sigma}=G_0|T_{\sigma}|^2,
\label{cond}
\end{equation}
where $T_{\sigma}$ is the total transmission coefficient and $G_0=e^2/h^2$. To calculate the transmission probability for each spin channel we need to solve the time dependent Schrodinger equation which reads
\begin{equation}
[H(t)-i\partial_t]|\Psi(t)\rangle=0.
\label{eqtds}
\end{equation}
Using the Floquet formalism  for a time periodic Hamiltonian of the form $H(t)=H_0+2H_1\cos(\omega t)$,\cite{reichl} we can write the steady state solution in the following form $|\Psi(t)\rangle=e^{-i\varepsilon t}|\Phi(t)\rangle$ where $|\Phi(t)\rangle$ satisfies the following equation
\begin{equation}
[H(t)-i\partial_t]|\Phi(t)\rangle=\varepsilon|\Phi(t)\rangle.
\label{eqtds1}
\end{equation}
where $|\Phi(t)\rangle=|\Phi(t+2\pi/\omega)\rangle$ is a periodic function of time and $\varepsilon$ is the quasienergy or Floquet energy. Using the following solution for Eq. (\ref{eqtds1})
\begin{equation}
|\Phi(t)\rangle=\sum_{n=-\infty}^{\infty}e^{-in\omega t}|\phi_n\rangle,
\label{flo}
\end{equation} 
the eigenvalue equation given in (\ref{eqtds1}) becomes
\begin{equation}
H_0|\Phi_n\rangle+H_1\left(|\Phi_{n+1}\rangle+|\Phi_{n-1}\rangle\right)=(\varepsilon+n\omega)|\Phi_{n}\rangle.
\label{flo1}
\end{equation}
The stationary state of the total Hamiltonian can be written as \cite{ore}
\begin{equation}
|\Phi_n^{\sigma}\rangle=\sum_{j\neq 0}\phi_{j,n}^{\sigma}c_{j,\sigma}^{\dagger}|0\rangle|s\rangle+\phi_{0,n}^{\sigma}c_{0,\sigma}^{\dagger}|0\rangle|s+\sigma\rangle,
\label{eqstatio}
\end{equation}
where $|0\rangle$ is the vacuum state, $|s\rangle$ is the spin state of the SMM and $\phi_{j,n}^{\sigma}$ is the probability amplitude to find the electron at site $j$ with quasienergy $\varepsilon$ and spin $\sigma$. Inserting eq. (\ref{eqstatio}) into eq. (\ref{flo1}) results in the following recursion relation for the amplitudes $\phi_{j,n}^{\sigma}$ :
\begin{widetext}
\begin{eqnarray}
\label{diffeq}
\mbox{for $|j|>1$ we have} \\ \nonumber
\varepsilon_{n,\sigma}\phi_{j,n}^{\sigma}&=&-J(\phi_{j+1,n}^{\sigma}+\phi_{j-1,n}^{\sigma}),\\ \nonumber
\mbox{for $j=0,1,-1$ we have} \\ \nonumber
\tilde{\varepsilon}_{n,\sigma}\phi_{0,n}^{\sigma}&=&-J^{\prime}(\phi_{1,n}^{\sigma}+\phi_{-1,n}^{\sigma})-\frac{V_g}{2}(\phi_{0,n+1}^{\sigma}+\phi_{0,n-1}^{\sigma}),\\  \nonumber
\varepsilon_{n,\sigma}\phi_{1,n}^{\sigma}&=&-J\phi_{2,n}^{\sigma}-J^{\prime}\phi_{0,n}^{\sigma}, \\ \nonumber
\varepsilon_{n,\sigma}\phi_{-1,n}^{\sigma}&=&-J\phi_{-2,n}^{\sigma}-J^{\prime}\phi_{0,n}^{\sigma}
\end{eqnarray}
\end{widetext}
where $\varepsilon_{n,\sigma}=\epsilon_{\sigma}+n\omega-\varepsilon_{SMM}-\epsilon_{OL}-J_{\parallel}s\delta_{\sigma}$ and $\tilde{\varepsilon}_{n\sigma}=\epsilon_{\sigma}+n\omega-\varepsilon_{SMM}^{\prime}-\epsilon_{OL}-J_{\parallel}s^{\prime}\delta_{\sigma}$, where $\delta_{\sigma}=+1(-1)$ for $\sigma=\uparrow(\downarrow)$, and $\varepsilon_{SMM}$ and $\varepsilon_{SMM}^{\prime}$ denotes the energy of the uncharged or singly charged SMM, respectively.  \\
For a incoming electron with spin $\sigma$, wave number $k_0$ for the mode $n=0$ and with Floquet energy $\varepsilon_{0,\sigma}=-2J\cos(k_0)$, the solution of the above equations for all the channels $n$ can be written as
\begin{eqnarray} 
\label{eqwave}
|\Phi_n^{\sigma}\rangle=\sum_{j<0}[\delta_{n,0}Ae^{ik_0j}c_{j,\sigma}^{\dagger}+e^{-ik_nj}r_{n,\sigma}c_{j,\sigma}^{\dagger}]|0\rangle|s\rangle, \\ \nonumber
+\sum_{j>0}e^{ik_nj}t_{n,\sigma}c_{j,\sigma}^{\dagger}|0\rangle|s\rangle+E_{n,\sigma}\frac{J}{J^{\prime}}c_{0,\sigma}^{\dagger}|0\rangle|s+\sigma\rangle.
\end{eqnarray}
Substituting Eq. (\ref{eqwave}) into Eq. (\ref{diffeq}) we get the dispersion relation $\varepsilon_{n,\sigma}=-2J\cos(k_n)$and we obtain the following current conservation equation
\begin{equation}
E_{n,\sigma}=t_{n,\sigma}=r_{n,\sigma}+\delta_{n,0}A.
\label{ccur}
\end{equation}
Using Eq. (\ref{ccur}) we can get the amplitudes at each site for the Floquet modes, which are
\begin{equation}
\phi_{j,n}^{\sigma}=E_{n,\sigma}e^{ik_n|j|}+\delta_{j,0}E_{n,\sigma}\left(\frac{J}{J^{\prime}}-1\right)+2i\delta_{n,0}\sin(k_0j)\theta(-j)
\label{amp}
\end{equation}
where $\theta(-j)$ is the Heaviside function. Plugging Eq. (\ref{amp}) in to Eq. (\ref{diffeq}) we get the following recursive relation for the coefficientes $E_{n,\sigma}$
\begin{eqnarray}
\label{trans}
E_{n+1,\sigma}+E_{n-1,\sigma}&=&-\frac{2\tilde{\varepsilon}_{n\sigma}}{V_g}E_{n,\sigma} \\ \nonumber
& & -\frac{4(J^{\prime})^2}{V_gJ}\left[E_{n,\sigma}e^{ik_n}-iA\delta_{n,0}\sin(k_n)\right] 
\end{eqnarray}
Equation (\ref{trans}) is a recurrence relation that needs to be solved by requiring the convergence $E_{|n|\rightarrow\infty}\rightarrow 0$.\\
The total transmission probability for the current across the SMM is calculated in terms of $E_{n,\sigma}$, which gives
\begin{widetext}
\begin{equation}
\label{eqlan}
T_{\sigma}(\epsilon_{\sigma},B_x)=Re\left[\frac{E_{0,\sigma}}{A}\right]=Re\left[\frac{u_{k}(1-\beta)}{(1-\beta)u_{k}-i\beta\varepsilon_{0,\sigma}+i(\varepsilon_{SMM}^{\prime}+\epsilon_{OL}+J_{\parallel}s^{\prime}\delta_{\sigma})-i\frac{V_g}{2}\left(\frac{E_{n+1,\sigma}}{E_{0,\sigma}}+\frac{E_{n-1,\sigma}}{E_{0,\sigma}}\right)}\right],
\end{equation}
\end{widetext}
where $u_{k}=2J\sin(k_0)$ is the velocity of the incoming electron and $\beta=1-(J^{\prime}/J)^2$. To calculate the ratios $E_{n\pm 1,\sigma}/E_{0,\sigma}$ we need to solve the recursive relation given in Eq. (\ref{trans}) numerically for a given set of physical parameters. \\
It is interesting to consider the zero transmission resonances that occur when $T_{\sigma}=0$ at certain energies $\epsilon_{\sigma}$. We can derive and approximation to estimate the location of the zero transmission resonances for small $V_g$ using Eq. (\ref{trans}) for $n\neq 0$, which can be written as $E_{n+1,\sigma}+E_{n-1,\sigma}=\gamma E_{n,\sigma}$
where
\begin{align}
\gamma=\frac{2}{V_g}\left[(\varepsilon_{SMM}^{\prime}+\epsilon_{OL}+J_{\parallel}s^{\prime}\delta_{\sigma})-\beta(\epsilon_{\sigma}+n\omega)\right.\nonumber \\ \left.-sgn(n)(1-\beta)\sqrt{(\epsilon_{\sigma}+n\omega)^2-4J^2}\right]
\label{res}
\end{align}
For small driving amplitudes, i.e. $V_g\rightarrow 0$, the points of zero transmission are given by
\begin{align}
\epsilon_{\sigma}=\mp\omega-\frac{\beta(\varepsilon_{SMM}^{\prime}+\epsilon_{OL}+J_{\parallel}s^{\prime}\delta_{\sigma})}{1-2\beta}\pm\frac{(1-\beta)}{1-2\beta}\nonumber \\ \times\sqrt{(\varepsilon_{SMM}^{\prime}+\epsilon_{OL}+J_{\parallel}s^{\prime}\delta_{\sigma})^2+4J^2(1-2\beta)}.
\label{ztrans}
\end{align}
For any given energy $\epsilon_{\sigma}$ it is possible to find a frequency $\omega$ so that there are always zero transmission points given by Eq. (\ref{ztrans}).\cite{eggert0} Equation (\ref{ztrans}) is valid for $J^{\prime}\geq J$ ($\beta\leq 0$), i.e. the strong coupling regime. In Figure (\ref{oscillate}) we show the zero transmission points as a function of the local transverse magnetic field for a SMM with spin $s=2$. Interestingly, the zero transmission points oscillate as a function of the transverse magnetic field. 
In Figure (\ref{tbz}) we show the total transmission probability of the current through the SMM transistor for given values of the transverse magnetic field and for the case when $\beta=0$. The resonances shown in Fig. (\ref{tbz}) are related to the phenomenon of Fano resonances. This interference effect arises when one or more bound states interact with the continuum states, leading to an asymmetric lineshape.
We are now ready to plot the conductance as a function of the electron energy and transverse magnetic field using Eq. (\ref{eqlan}). Figure (\ref{conduc}) and Figure (\ref{conduc1}) shows how the conductance oscillates as a function of the transverse magnetic field due to the Berry phase interference between different spin tunneling paths for a SMM with spin $s=2$ for the case when $\beta=0$ and $\beta=-0.5$, respectively. This demonstrates that it is possible to detect the Berry phase interference in a SMM transistor connected to unpolarized leads by using a local combination of a periodically driven gate voltage and a transverse magnetic field. Note how the amplitudes of the oscillations in the conductance decrease when we increase the coupling strength between the leads and the SMM. \\
\\
It is interesting to calculate the transmission at very low frequencies which means that for this case we can take the electron´s velocity approximately the same, i.e. $k_n\approx k_0$. In this limit we can obtain an approximate analytic expression for the transmission by using $E_{n,\sigma}=E_{0,\sigma}\lambda^{\pm n}$ in Eq. (\ref{trans}), which gives
\begin{equation}
\lambda^2-\frac{4J}{V_g}\left(\beta e^{ik_0}-i\sin(k_0)+\frac{\varepsilon_{SMM}^{\prime}+\epsilon_{OL}+J_{\parallel}s^{\prime}\delta_{\sigma}}{2J}\right)\lambda+1=0.
\label{trans0}
\end{equation}
Solving Eq. (\ref{trans0}) for $\lambda$ we have for low frequencies that
\begin{widetext}
\begin{equation}
\label{low}
\frac{E_{\pm 1,\sigma}}{E_{0,\sigma}}\approx\frac{2J}{V_g}\left(\beta e^{ik_0}-i\sin(k_0)+\frac{\varepsilon_{SMM}^{\prime}+\epsilon_{OL}+J_{\parallel}s^{\prime}\delta_{\sigma}}{2J}\right)\pm\sqrt{\left(\frac{2J}{V_g}\left(\beta e^{ik_0}-i\sin(k_0)+\frac{\varepsilon_{SMM}^{\prime}+\epsilon_{OL}+J_{\parallel}s^{\prime}\delta_{\sigma}}{2J}\right)\right)^2-1}.
\end{equation}
\end{widetext}
Substituting Eq. (\ref{low}) into Eq. (\ref{eqlan}) we obtain an analytic expression for the transmission for low frequencies which is given by
\begin{equation}
T_{\sigma}(\epsilon_{\sigma})=Re\left[\frac{u_{k_0}(1-\beta)}{(1-\beta)u_{k_0}-i\beta\epsilon_{\sigma}-i2J\left(\beta e^{ik_0}-i\sin(k_0)\right)}\right].
\label{low1}
\end{equation}
Note that the transmission for low frequencies, i.e. $\omega\rightarrow 0$, does not depends on the energy of the molecule which implies that we do not have oscillations as a function of the transverse magnetic field in agreement with Ref. \cite{gabriel1}.\\
In conclusion, we have investigated electron transport through a single molecule magnet transistor with a local periodically driven gate voltage and transverse magnetic field. We have shown that the conductance through the SMM transistor presents Berry phase oscillations as a function of the transverse magnetic field. This results provide a new way to observe the interference between quantum spin tunneling paths using a local periodically driven gate voltage and a transverse magnetic field using unpolarized leads in contrast with the method presented in Ref. \cite{gabriel1} where they show that for a constant gate voltage it is necessary to have polarized leads in opposite directions in order to detect the Berry phase oscillations in the current. 
\\ \\
I would like to acknowledge support by the program ``C\'atedras CONACYT" and from project 105 of ``Centro Mexicano de Innovaci\'on en Energ\'ia Solar" and by the National Laboratory program from CONACYT through the Terahertz Science and Technology National Lab (LANCYTT). I would like to thank S. Eggert and M.N. Leuenberger for useful discussions at the beginning of this work.

\end{document}